\documentclass[fleqn,usenatbib]{mnras}

\usepackage{newtxtext,newtxmath}

\usepackage[T1]{fontenc}

\DeclareRobustCommand{\VAN}[3]{#2}
\let\VANthebibliography\thebibliography
\def\thebibliography{\DeclareRobustCommand{\VAN}[3]{##3}\VANthebibliography}

\usepackage{graphicx}	
\usepackage{amsmath}	

\usepackage{siunitx}  
\DeclareSIUnit\Mpc{Mpc}
\DeclareSIUnit\Gpc{Gpc}
\DeclareSIUnit\solarmass{\ensuremath{\mathit{M}_{\sun}}}
\DeclareSIUnit\h{\ensuremath{\mathit{h}}}

\newcommand{\ezmock}{\textsc{ezmock}}
\newcommand{\ezmocks}{\textsc{ezmocks}}
\newcommand{\fastpm}{{\textsc{fastpm}}}
\newcommand{\cola}{\textsc{cola}}

\newcommand{\bsm}{\boldsymbol}

\title[Covariance matrices for variance-suppressed sims]{Covariance matrices for variance-suppressed simulations}

\author[Tony Zhang et al.]{%
Tony Zhang,$^{1}$\thanks{E-mail: txz@stanford.edu}
Chia-Hsun Chuang,$^{2,3}$\thanks{E-mail: chuangch@stanford.edu}
Risa H. Wechsler,$^{1,2,4}$
Shadab Alam,$^{5}$
\newauthor
Joseph DeRose,$^{6}$
Yu Feng,$^{7}$
Francisco-Shu Kitaura,$^{8,9}$
Marcos Pellejero-Ibanez,$^{10}$
\newauthor
Sergio~Rodr\'iguez-Torres,$^{11}$
Chun-Hao To,$^{12}$
Gustavo Yepes,$^{11,13}$
Cheng Zhao,$^{14}$
\\
$^{1}$Department of Physics, Stanford University, 382 Via Pueblo Mall, Stanford, CA 94305, USA\\
$^{2}$Kavli Institute for Particle Astrophysics and Cosmology, Stanford University, 452 Lomita Mall, Stanford, CA 94305, USA\\
$^{3}$Department of Physics and Astronomy, University of Utah, Salt Lake City, UT 84112, USA\\
$^{4}$SLAC National Accelerator Laboratory, 2575 Sand Hill Road, Menlo Park, CA 94025, USA\\
$^{5}$Institute for Astronomy, University of Edinburgh, Royal Observatory, Blackford Hill, Edinburgh EH9 3HJ, UK\\
$^{6}$Physics Division, Lawrence Berkeley National Laboratory, 1 Cyclotron Road, Berkeley, CA 94720, USA\\
$^{7}$Berkeley Center for Cosmological Physics, Department of Physics, University of California Berkeley, Berkeley, CA 94720, USA\\
$^{8}$Instituto de Astrof\'{\i}sica de Canarias (IAC), C/V\'{\i}a L\'actea, s/n, E-38200 La Laguna, Tenerife, Spain \\
$^{9}$Departamento Astrof\'{\i}sica, Universidad de La Laguna (ULL), E-38206 La Laguna, Tenerife, Spain \\
$^{10}$ Donostia International Physics Center (DIPC), Paseo Manuel de Lardizabal, 4, E-20018 Donostia-San Sebasti\'an, Spain\\
$^{11}$Departamento de F\'isica Te\'{o}rica, M\'{o}dulo 8, Facultad de Ciencias, Universidad Aut\'{o}noma de Madrid, E-28049 Madrid, Spain\\
$^{12}$Center for Cosmology and AstroParticle Physics, Ohio State University, Columbus, OH 43210, USA\\
$^{13}$CIAFF, Facultad de Ciencias, Universidad Aut\'{o}noma de Madrid, E-28049 Madrid, Spain\\
$^{14}$Institute of Physics, Laboratory of Astrophysics, Ecole Polytechnique F\'ed\'erale de Lausanne (EPFL), Observatoire de Sauverny, CH-1290 Versoix, Switzerland\\
}

\date{Accepted XXX. Received YYY; in original form ZZZ}

\pubyear{2021}

\begin{document}
\label{firstpage}
\pagerange{\pageref{firstpage}--\pageref{lastpage}}
\maketitle

\begin{abstract}
Cosmological $N$-body simulations provide numerical predictions of the structure of the Universe
against which to compare data from ongoing and future surveys,
but the growing volume of the Universe mapped by surveys
requires correspondingly lower statistical uncertainties in simulations,
usually achieved by increasing simulation sizes
at the expense of computational power.
It was recently proposed to reduce simulation variance
without incurring additional computational costs
by adopting fixed-amplitude initial conditions.
This method has been demonstrated not to introduce bias in various statistics,
including the two-point statistics of galaxy samples
typically used for extracting cosmological parameters from galaxy redshift survey data,
but requires us to revisit current methods
for estimating covariance matrices of clustering statistics for simulations.
In this work, we find that it is not trivial to construct covariance matrices analytically for fixed-amplitude simulations,
but we demonstrate that \ezmock{}
(Effective Zel'dovich approximation mock catalogue),
the most efficient method for constructing mock catalogues with accurate two- and three-point statistics,
provides reasonable covariance matrix estimates for such simulations.
We further examine
how the variance suppression obtained by amplitude-fixing
depends on three-point clustering, small-scale clustering, and galaxy bias,
and propose intuitive explanations for the effects we observe
based on the \ezmock{} bias model.
\end{abstract}




\section{Introduction}

The study of the large-scale structure of the Universe
has become a precision science
in recent years,
thanks to such surveys as the photometric DES\footnote{\url{http://www.darkenergysurvey.org}}
(Dark Energy Survey, \citealt{2005astro.ph.10346T})
and the spectroscopic SDSS%
\footnote{\url{http://www.sdss.org/sdss-surveys}}
(Sloan Digital Sky Survey, \citealt{2017AJ....154...28B}).
The total volume of the Universe mapped with galaxy surveys will continue to increase
as the next generation of large ground- and space-based experiments comes online,
including DESI\footnote{\url{http://desi.lbl.gov/}} (Dark Energy Spectroscopic Instrument; \citealt{Schlegel:2011zz,Levi:2013gra}),
4MOST\footnote{\url{http://www.4most.eu/}} (4-metre Multi-Object Spectroscopic Telescope; \citealt{deJong:2012nj}),
HETDEX\footnote{\url{http://hetdex.org}} (Hobby--Eberly Telescope Dark Energy Experiment; \citealt{Hill:2008mv}),
J-PAS\footnote{\url{http://j-pas.org}} (Javalambre Physics of the Accelerating Universe Astrophysical Survey; \citealt{Benitez:2014ibt}),
PFS\footnote{\url{https://pfs.ipmu.jp}}(Subaru Prime Focus Spectrograph; \citealt{2014PASJ...66R...1T}),
LSST\footnote{\url{http://www.lsst.org/lsst/}} (Legacy Survey of Space and Time; \citealt{Abell:2009aa}), 
Euclid\footnote{\url{http://www.euclid-ec.org}} \citep{Laureijs:2011gra}, 
and the Roman Space Telescope\footnote{\url{https://roman.gsfc.nasa.gov}} \citep{Spergel:2013tha}.

Theoretical modelling of galaxy clustering,
including galaxy bias and peculiar motions,
is crucial for testing cosmological models against observations.
On large scales, these theoretical models are often based on perturbation theory with approximations,
and their validation requires fully non-linear numerical solutions,
generally in the form of $N$-body simulations tracking the full growth of structure over the Universe's history.
These simulations must be sufficiently large compared to the volumes sampled in surveys
while maintaining sufficient mass resolution
to resolve the dark matter haloes that host the galaxies
typically detected in surveys.
As survey volumes continue to grow,
so too must the size of these $N$-body simulations.

These dual requirements of simulation volume and mass resolution
are difficult to meet with current computational power.
Indeed, a single simulation with the required halo mass resolution
($\sim\SI{1e11}{\solarmass\per\h}$, \citealt{Gonzalez-Perez:2017mvf})
covering the entire survey volume of DESI [$\sim \SI{70}{(\Gpc.\h^{-1})^3}$]
would demand an enormous number of particles
($\gtrsim\num{16000}^3$ in a box of side length \SI{4}{\Gpc\per\h}).
However, the largest $N$-body simulations to date, e.g.
MillenniumXXL  \citep{Angulo:2012ep},
MICE \citep{Fosalba:2013wxa},
MultiDark \citep{Klypin:2014kpa},
Dark Sky \citep{Skillman:2014qca}, 
OuterRim \citep{Habib:2014uxa},
FLAGSHIP \citep{Potter:2017aa},
UNIT \citep{Chuang:2018ega},
Uchuu \citep{Ishiyama:2020vao},
and AbacusSummit \citep{Maksimova:2021ynf},
remain well below the particle numbers we require.
This problem is compounded
by the need to reduce the uncertainty in these simulations
to levels well below the statistical uncertainty of observations
in order to maximize the information extracted from survey data:
a simulation with size merely \emph{equal} to the survey volume is insufficient.

\citet{Angulo:2016hjd} proposed to suppress the variance in these simulations
by removing amplitude fluctuations
in the various $k$-modes in the initial conditions of a simulation.
Such fixed-amplitude initial conditions
have been employed in hydrodynamical simulations \citep{Villaescusa-Navarro:2018bpd}
and have been tested on Lyman-$\alpha$ forest statistics \citep{Anderson:2018zkm}.
\citet{Chuang:2018ega} also tested the method on the clustering measurements of galaxy redshift surveys.
All found that the variance-suppression method
provides more precise predictions without introducing bias,
enabling simulations with large effective volume
at much lower computational cost.
However, the reduced variance in these simulations still must be quantified
for comparison with the variance in survey data
(whose estimation is an important problem in its own right),
and we must verify that the methods typically employed
to estimate covariance matrices for the clustering statistics
of galaxy catalogues
can provide reliable estimates
when those catalogues are derived from variance-suppressed simulations.

Traditionally, covariance matrix estimation
for both observation and simulation data
has been done with mock galaxy catalogues.
Various methods exist for generating such mock catalogues,
but we broadly categorize them into two classes
based on how they construct their halo catalogues.
The first class defines haloes by applying a halo finder
on simulated dark matter particles,
and includes methods such as
\textsc{pthalos} \citep{Manera:2012sc, Manera:2014cpa},
\textsc{pinocchio} \citep[PINpointing Orbit-Crossing Collapsed Hierarchical Objects,][]{Monaco:2001jg,Monaco:2013qta},
\textsc{peak patch} \citep{Bond:1993we},
\fastpm{} \citep{Feng:2016yqz},
\cola{} (COmoving Lagrangian Acceleration simulation; \citealt{Tassev:2013pn,Izard:2015dja}),
and \textsc{glam} \citep{Klypin:2017iwu}.
These methods tend to be memory-intensive,
as they require a large number of particles
in order to resolve haloes;
indeed, \fastpm{}, \cola{}, and \textsc{glam} are essentially efficient $N$-body simulations.
As a consequence, however,
mock catalogues produced with such codes using fixed-amplitude initial conditions
naturally yield good covariance matrix estimates for fixed-amplitude simulations.

The second class of methods populates haloes based on bias models
using coarse-resolution dark matter density fields.
In these models, halo creation is stochastic,
but is calibrated to reproduce the clustering statistics
of a reference catalogue
(obtained, for example, from survey data or from a higher resolution simulation).
This second class includes such methods as
\textsc{log-normal} \citep{Coles:1991if},
\textsc{patchy} (PerturbAtion Theory Catalog generator of Halo and galaxY distributions; \citealt{Kitaura:2013cwa,Kitaura:2014mja}),
\textsc{halogen} \citep{Avila:2014nia},
\textsc{qpm} (quick particle mesh; \citealt{White:2013psd}),
\textsc{bam} (Bias Assignment Method to generate mock catalogues; \citealt{Balaguera-Antolinez:2018cuq}),
and \ezmock{} (Effective Zel'dovich approximation mock catalogue; \citealt{Chuang:2014vfa}).

The computational cost of this second class of methods is much lower,
but the validity of their use in covariance matrix estimation is less obvious.
Some past work exists validating these methods for covariance matrix estimation
and comparing them with each other and with those of the first class
(see e.g. \citealt{Chuang:2014toa,Blot:2018oxk,Colavincenzo:2018cgf,Lippich:2018wrx}),
but more study is desired to fully validate their use in upcoming surveys,
especially for covariance matrix estimation of observational data.

Still, these methods have gained traction
for their key advantage of efficiency.
\ezmock{} in particular delivers reasonable accuracy
at very low computational cost,
and indeed, was employed extensively in the cosmological analysis
of the final eBOSS galaxy sample
(see e.g.
\citealt{
    Avila:2020rmp,
    Bautista:2020ahg,
    Gil-Marin:2020bct,
    Hou:2020rse,
    Kong:2020nhv,
    Mohammad:2020kwk,
    Neveux:2020voa,
    Raichoor:2020vio,
    Ross:2020lqz,
    Tamone:2020qrl,
    deMattia:2020fkb,
    Alam:2020sor,
    Zhao:2020bib,
    Zhao:2020tis%
}.
It remains unclear, however,
whether \ezmock{}, or any method in this second class,
can be extended to estimate covariance matrices of variance-suppressed simulations.

In this study,
we show that \ezmock{} can provide reasonable covariance matrix estimates for such simulations.
The rest of this paper is organized as follows.
We outline the simulations used in this study in \autoref{sec:sim},
then present our findings in \autoref{sec:results}.
In particular, we show that fixed-amplitude \ezmock{} can reproduce
the covariance matrices of a reference fixed-amplitude simulation
in \autoref{sec:compare},
then proceed in \autoref{sec:effects} to examine in turn
the effects of three-point clustering, small-scale clustering, and galaxy bias
on the level of variance suppression obtained by fixing amplitudes,
showing in the process how \ezmock{} can provide intuitive explanations for our observations.
We summarize and conclude in \autoref{sec:conclusion}.

\section{Simulations}
\label{sec:sim}

\subsection{\fastpm{} catalogues}

In order to study how \ezmock{} reproduces the covariance matrix of a reference simulation,
we require a large number of reference simulations whose covariance matrices we will estimate.
To this end, we use a set of simulations produced with the \fastpm{} code \citep{Feng:2016yqz}.
The \fastpm{} code relies on accelerated particle--mesh solvers,
which have recently been shown to produce accurate halo populations
(compared to full $N$-body calculations)
when enhanced with various techniques
(cf.\@ the \cola{} code; \citealt{Tassev:2013pn}).
\fastpm{} employs a pencil domain-decomposition Poisson solver
and a Fourier-space four-point differential kernel to compute the force.
Given that \fastpm{} simulations are essentially $N$-body simulations,
we use them as our reference in this study.

Our simulations, 200 in all, are a subset of the \fastpm{} simulations created by the UNIT project
to test the behaviour of the variance-suppression method \citep{Chuang:2018ega}.
All of these simulations are publicly available.\footnote{\url{http://www.unitsims.org}}
Of the 200 simulations we use here,
half were produced with fixed-amplitude initial conditions and half without.
All are at $z=1$.
Each simulation was generated with $1024^3$ particles in a box \SI{1}{\Gpc\per\h} on each side
and simulated for 100 time-steps.
A Friends-of-Friends halo finder was used to identify haloes;
the minimum halo mass is \SI{1.68E12}{\solarmass\per\h}.

Throughout this work we use the 
same cosmological parameters as these \fastpm{} simulations:
$\Omega_{\mathrm{m}}=0.3089$,
$h\equiv H_0/(\SI{100}{\km\per\second\per\Mpc}) = 0.6774$,
$n_s=0.9667$,
and $\sigma_8=0.8147$
\citep[based on][table 4]{2016A&A...594A..13P}.

\subsection{\ezmock{}}
\label{sec:ezmock}

\ezmock{} \citep{Chuang:2014vfa}
is a method for generating mock catalogues
based on application of a bias model
to a coarse-resolution dark matter field.
The dark matter field is constructed
from the Zel'dovich approximation (ZA) density field.
\ezmock{} absorbs non-linear effects and halo bias
(i.e.\@ linear, non-linear,
deterministic, and stochastic bias)
into an effective model with only a few free parameters,
which can be efficiently calibrated with $N$-body simulations.
We use the slightly modified version described in \citet{Baumgarten:2018jcn},
consisting of the following three steps:

\begin{enumerate}
    \item \textbf{Generate the dark matter field.}
    In the Lagrangian formulation of cosmological fluid dynamics,
    we describe the motion of a particle originally at $\boldsymbol q$
    to a position $\boldsymbol x$ at cosmic time $t$
    by a Lagrangian displacement field $\bsm\Psi$:
    \begin{equation}
        \boldsymbol x(\boldsymbol q, t) = \boldsymbol q + \boldsymbol\Psi(\boldsymbol q, t).
    \end{equation}
    The first-order Lagrangian perturbation theory solution
    to the equations of motion is given by the ZA
    \begin{equation}
        \bsm\Psi(\bsm q) = \int \frac{d^3 \bsm k}{(2\upi)^3} \mathrm{e}^{i\bsm k\cdot\bsm q} \frac{i\bsm k}{k^2} \hat\delta(\bsm k)
    \end{equation}
    where $\hat\delta(\bsm k)$ is the fractional density perturbation in Fourier space
    (for a review, see e.g. \citealt{Bernardeau:2001qr}).
    
    We construct this displacement field in the ZA
    and extrapolate to the redshift of the reference halo or galaxy sample
    (in our case, $z=1$).
    To generate our dark matter density field,
    we initialize a uniform square lattice of dark matter particles
    and evolve the particles according to $\bsm\Psi$.
    We can apply fixed-amplitude initial conditions in this step
    by selecting the $\hat\delta(\bsm k)$ to have random phases but predetermined fixed amplitudes
    rather than drawing them from Gaussian distributions.
    
    \item
    \label{item:ezmock-pdf-mapping}
    \textbf{Determine final object densities
    using the dark matter density
    and the halo probability distribution function (PDF).}
    We model the PDF of the final mock catalogue,
    the number of grid cells containing $n$ objects as a function of $n$,
    as
    \begin{equation}
        \label{eq:pdf}
        P(n) \propto A^n,
    \end{equation}
    normalized to give the desired final object number density (a free parameter).
    Greater dark matter density should yield greater object density.
    Thus, we might naively sort the grid cells $\bsm r$
    by increasing dark matter density $\rho_0(\bsm r)$
    and assign each cell the number of objects it should contain:
    the first $P(0)$ will contain no objects, the next $P(1)$ will contain one, etc.
    
    In practice, we use modified ``densities'' $\rho_s(\bsm r)$
    for this rank-ordering procedure
    to allow more tuning of our clustering.
    Beginning with dark matter densities $\rho_0(\bsm r)$ in each cell
    obtained using the cloud-in-cells (CIC) particle assignment scheme \citep[see e.g.][]{Hockney1981},
    cells with densities below some density cut $\rho_c$ are assigned $\rho_s = 0$.
    Adjusting $\rho_c$ allows us to modify the bispectrum of the catalogues.
    
    On the other hand, if $\rho_0 \ge \rho_c$, we compute
    \begin{equation}
        \rho_s(\bsm r) =
        (1 - \mathrm{e}^{-\rho_0(\bsm r)/\rho_a}) \begin{cases}
          1 + G(\bsm r) & G(\bsm r) \ge 0 \\
          \mathrm{e}^{G(\bsm r)} & G(\bsm r) < 0
        \end{cases}
    \end{equation}
    where, for each cell,
    $G(\bsm r)$ is independently drawn from a Gaussian distribution of some specified width $\lambda$
    [typically left fixed,
    having observed in the course of previous work
    \citep{Baumgarten:2018jcn}
    that its effect on clustering statistics
    was highly degenerate with that of the other parameters].
    This procedure effectively introduces some scatter into the PDF mapping procedure described above.
    The non-linear function $1 - \mathrm{e}^{-\rho_0/\rho_a}$
    introduces saturation behaviour
    at a characteristic density~$\rho_a$ (another free parameter),
    converging to 1 when $\rho_0 \gg \rho_a$.

    \item \textbf{Assign objects.}
    The mock catalogue is populated
    by selecting at random a subset of the original dark matter particles
    such that each cell contains the number of objects assigned to it
    in the previous step.
\end{enumerate}

We generate three sets of \ezmock{} catalogues,
each consisting of $2\times1000$ catalogues
of size $(\SI{1}{\Gpc\per\h})^3$
and number density $\overline{n} = \SI{2.343e6}{\per(\Gpc\per\h)\cubed}$
(chosen to match the \fastpm{} catalogues)
generated on a $256^3$ grid.
Within each set, half the catalogues are produced using normal initial conditions
and half using fixed-amplitude initial conditions.

The first set of \ezmocks{} is a fiducial set
with parameters chosen to reproduce the clustering statistics of the \fastpm{} catalogues
($A = 0.37$ and $\rho_a, \rho_c = 2.25, 0$ in units of objects per grid cell).
The second set has matching two-point clustering
but differs in its three-point clustering (i.e.\@ bispectrum)
($A = 0.37$ and $\rho_a, \rho_c = 0.75, 1$).
The third set maintains the linear-scale amplitude of the fiducial set
but differs in small-scale clustering
($A = 0.3$, $\rho_a, \rho_c = 3.1, 0$).
We fixed $\lambda = 10$.

\section{Results}
\label{sec:results}

\subsection{Covariance matrix estimation}
\label{sec:compare}

\begin{figure}
    \centering
    \includegraphics[width=0.9\columnwidth]{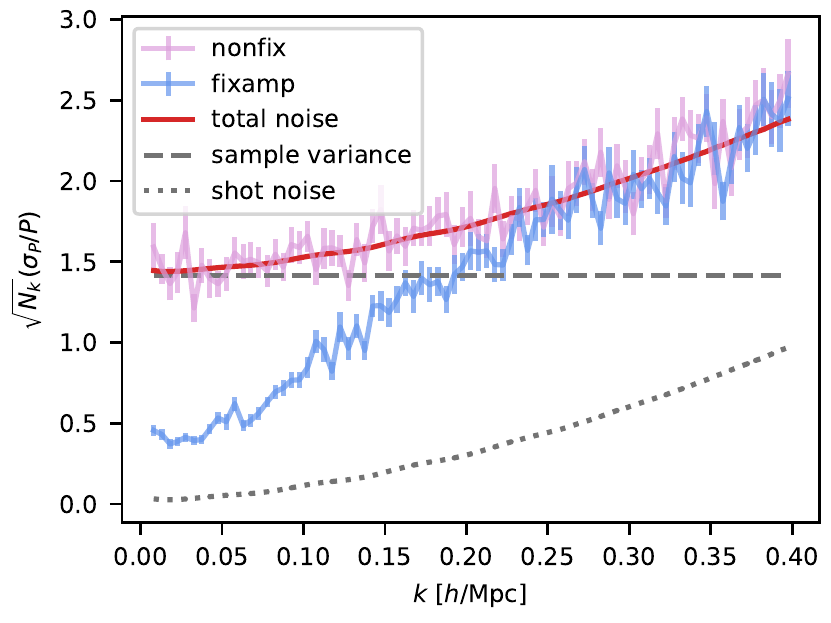}
    \caption{
    Comparison of fractional uncertainty in the power spectrum $P(k)$
    for the two sets of \fastpm{} catalogues,
    one produced with (blue; ``fixamp") and one produced without (pink; ``nonfix")
    fixed-amplitude initial conditions,
    against the theoretical fractional uncertainty (red).
    We scale by $\sqrt{N_k}$, where $N_k \sim k^2$ is the number of modes in each $k$-bin.
    This theoretical uncertainty is the sum of a Gaussian component (grey dashed line)
    and a Poisson shot-noise component (grey dotted line).
    The theoretical uncertainty agrees
    with that of the non-fixed-amplitude \fastpm{} simulations very well,
    but that of the fixed-amplitude \fastpm{} simulations
    can neither be described by the theoretical uncertainty
    nor its Gaussian or Poissonian components
    at small $k$.
    The computation of error bars is described in Appendix~\ref{sec:errorbars}.
    }
    \label{fig:gaussian_shotnoise}
\end{figure}

It is well known that the fractional uncertainty
in a band-averaged power spectrum value $P(k)$
for a Gaussian random field
under Poisson sampling
can be decomposed into a sum of a Gaussian sample variance
and a shot-noise contribution as
\begin{equation}
    \label{eq:pk-uncertainty}
    \frac{\sigma_{P(k)}}{P(k)} = \sqrt{\frac{2}{N_k}} \left( 1 + \frac{1}{\overline n P(k)} \right),
\end{equation}
where $N_k \sim k^2$ is the number of modes in the $k$-bin
and $\overline n$ is the mean number density
\citep[see][]{Feldman:1993ky}.
\autoref{fig:gaussian_shotnoise} confirms that this model
accurately accounts for the power spectrum uncertainty
in the non-fixed-amplitude \fastpm{} catalogues.

The fixed-amplitude \fastpm{} catalogues
show suppressed variance at linear scales (small $k$).
We might expect the residual uncertainty here
to be dominated by a shot-noise contribution,
as the initial dark matter power spectrum has no Gaussian sample variance (by construction),
but \autoref{fig:gaussian_shotnoise} shows that this is not the case.
The excess variance here is also too great to be easily explained
as the result of weak non-linear effects
affecting the evolution of the dark matter field;
\citet{Angulo:2016hjd} found
much lower dark matter power spectrum variance
in their fixed-amplitude simulations
at these $k$
(see the residuals for individual fixed-amplitude simulations
at $k < \SI{0.1}{\h\per\Mpc}$
in their fig.~2).
Since halo presence is intimately connected to
small-scale dark matter clustering,
the residual small-$k$ variance in the halo power spectrum
in the fixed-amplitude \fastpm{} simulations
may very well arise from large-$k$ dark matter power spectrum variance.
It is evidently non-trivial to build a theoretical model
for the covariance matrix of two-point statistics
of fixed-amplitude simulations,
even at linear scales.

\begin{figure*}
    \centering
    \includegraphics[width=2\columnwidth]{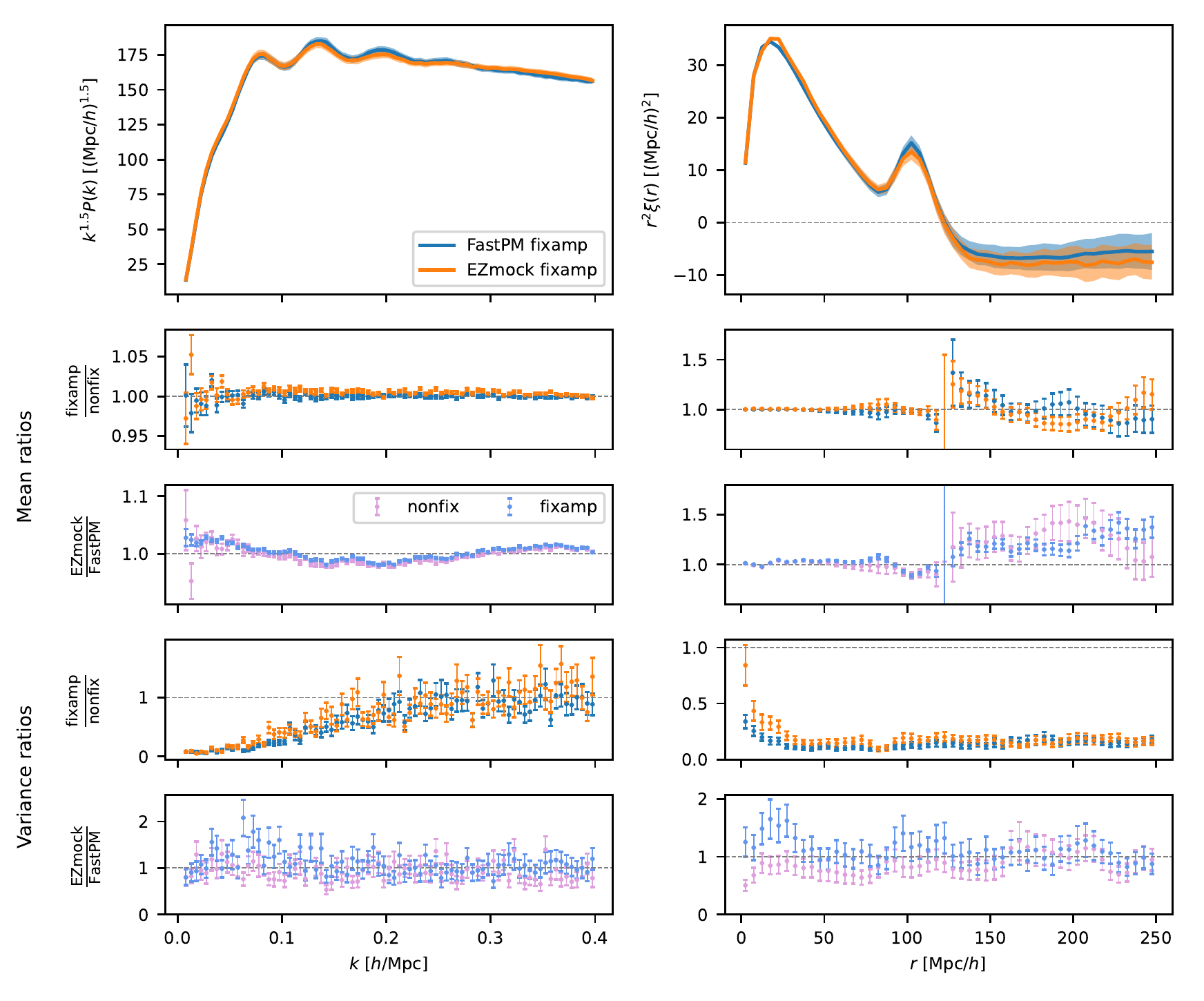}
    \caption{%
    Comparison of \fastpm{} catalogues and $2\times 100$ fitted \ezmocks{}.
    Top panels show the mean power spectrum~$P(k)$
    and correlation function~$\xi(r)$
    of the fixamp \fastpm{} and \ezmock{} catalogues
    (shaded bands show $\pm 1\sigma$).
    The next row,
    which shows 
    the ratio of the $P(k)$ and $\xi(r)$ means for both \fastpm{} and \ezmock{}
    for the fixamp catalogues
    and the corresponding means
    for the nonfix catalogues,
    indicates that the fixed-amplitude condition does not bias the mean values.
    The third row compares the means in a different way,
    plotting the ratio between the \ezmocks{} and the \fastpm{} catalogues for each amplitude condition
    (i.e.\@ nonfix \ezmock{} vs.\@ nonfix \fastpm{}
    and fixamp \ezmock{} vs.\@ fixamp \fastpm{}).
    We see that \ezmock{} reproduces the clustering measurements of the \fastpm{} catalogues.
    The final two rows are analogous to the second and third,
    but show ratios of variances.
    It is evident that the \ezmocks{}
    reproduce the variance suppression 
    seen in the \fastpm{} simulations,
    especially at large scales.
    Standard errors are computed per Appendix~\ref{sec:errorbars}.
    }
    \label{fig:fastpm-ezmock-compare}
\end{figure*}

Can a method like \ezmock{},
which produces mock catalogues based on a bias model,
reproduce the covariance matrix of these fixed-amplitude catalogues?
Yes.
In \autoref{fig:fastpm-ezmock-compare},
we compare the power spectrum and two-point correlation function
for a fiducial set of $2\times 100$ \ezmocks{} and the \fastpm{} catalogues.
The parameters are the same between the fixed-amplitude (``fixamp'')
and non-fixed-amplitude (``nonfix'') \ezmocks{}.
Thus, the second row shows that applying the fixed-amplitude condition
does not bias our clustering statistics,
consistent with previous studies \citep[see e.g.][]{Chuang:2018ega}.
In the bottom panels,
the variance suppression observed
in the power spectrum of the \fastpm{} boxes
is reproduced by the \ezmocks{}.
The variance suppression observed in the \fastpm{} boxes' correlation function measurements
is also reproduced reasonably at large scales,
though \ezmock{} underestimates the suppression at smaller scales.
This underestimation may be a limitation of the bias model underlying \ezmock{}
being ``too stochastic" at small scales,
as further tuning of fit parameters does not tend to improve this underestimation.

\begin{figure*}
  \centering
  \includegraphics[width=0.89\textwidth]{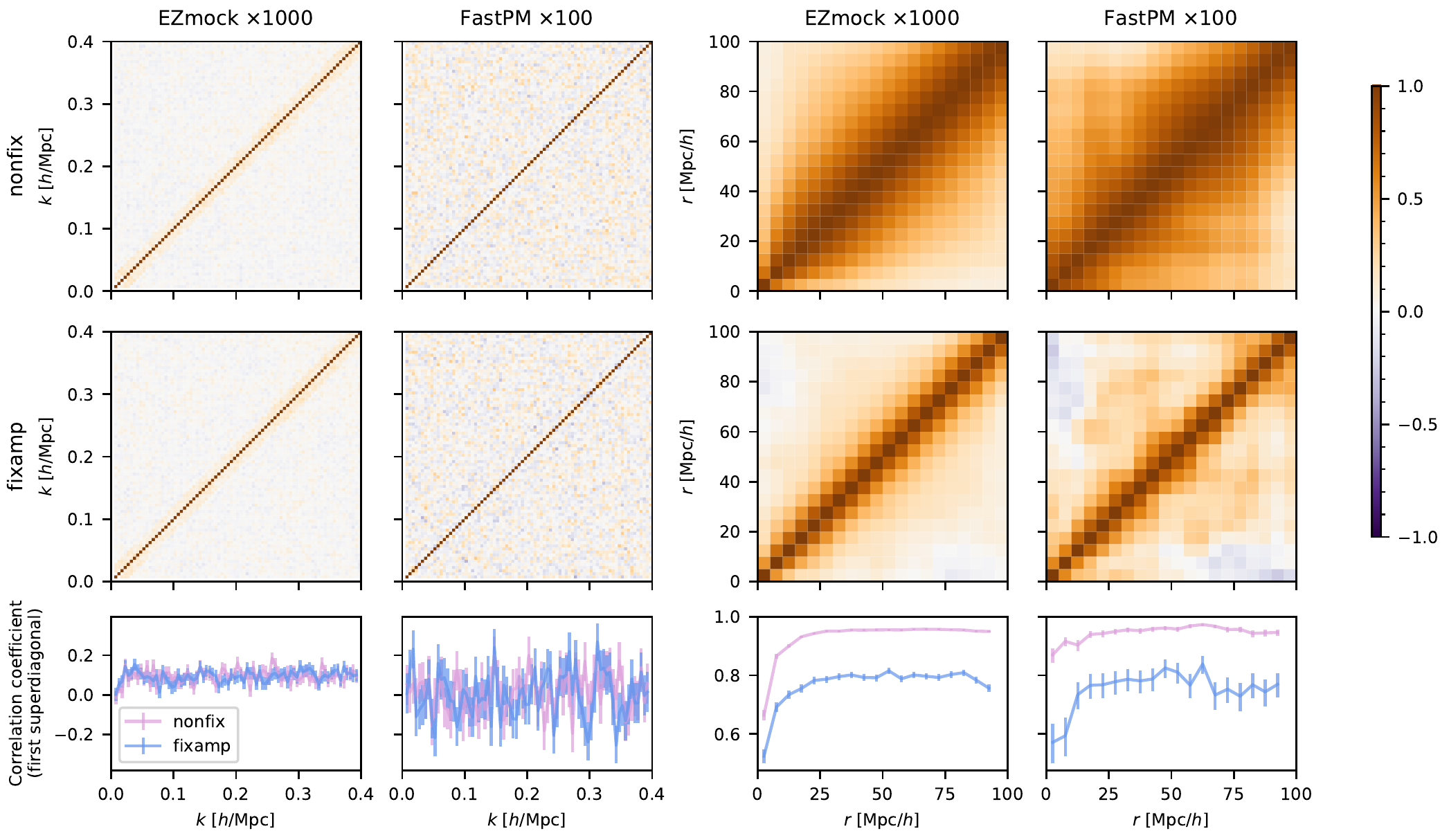}
  \caption{%
    Comparison of correlation matrices for the two-point clustering statistics of four sets of simulations:
    the ``nonfix" and the ``fixamp" boxes
    for both the fiducial \ezmocks{} and the reference \fastpm{} simulations.
    The \ezmock{} (\fastpm{}) sets consist of 1000 (resp.\@ 100) boxes each.
    The top two rows show
    correlation matrices for the nonfix and fixamp boxes.
    The correlation structure in the power spectrum
    is similar between the nonfix and fixamp boxes
    for both \ezmock{} and \fastpm{},
    but the fixamp correlation function values
    are less correlated than their nonfix counterparts,
    as confirmed by the lower panels,
    which show the first off-diagonal terms
    of the correlation matrices above
    [i.e.\@ correlations between $P(k)$ and $P(k + \Delta k)$ on the left
    and likewise with the correlation function on the right].
    We also find that the correlation structure
    between the corresponding sets
    of \ezmock{} and \fastpm{} boxes is quite similar,
    up to the statistical uncertainty of the \fastpm{} boxes.
    Standard errors are computed per Appendix~\ref{sec:errorbars}.
    }
    \label{fig:corr-summary}
\end{figure*}

We also compute normalized covariance matrices
(i.e.\@ correlation matrices)
of the power spectrum and correlation function
for the full set of fiducial \ezmocks{} (both fixed-amplitude and non-fixed-amplitude),
as well as for the reference \fastpm{} catalogues.
These correlation matrices are shown in \autoref{fig:corr-summary},
where we observe that for both the power spectrum and the correlation function,
the \ezmock{} correlation matrix agrees with that of our \fastpm{} boxes.

The stochastic bias we observe in the fixed-amplitude \ezmock{} catalogues
is introduced through the scattering procedure used in the PDF mapping,
as described in step~\ref{item:ezmock-pdf-mapping} of \autoref{sec:ezmock}.
The parameters involved are calibrated with the clustering measurements
of the reference catalogue (in this case, \fastpm{}).
We do \emph{not} tune additional parameters
to calibrate the covariance matrix separately.
That the covariance matrices obtained from the \ezmocks{}
agree with those of the \fastpm{} simulations
shows that the \ezmock{} bias model reasonably reproduces the stochastic bias
of the \fastpm{} haloes.

In \autoref{fig:corr-summary},
we further note that for both the \ezmocks{} and the \fastpm{} catalogues,
the power spectrum correlation matrices
are similar between the fixed-amplitude and non-fixed-amplitude simulations,
as underscored by the cuts plotting the first off-diagonal correlations:
fixed-amplitude initial conditions
do not bias the mode coupling in Fourier space.

In contrast, the correlation matrices for the correlation functions are dramatically different
between fixed-amplitude and non-fixed-amplitude catalogues.
This difference is not surprising, however.
Indeed, while one can compute the covariance matrix of the correlation function
from that of the power spectrum
by expressing the correlation function as a Fourier transform of the power spectrum,
the correlation matrix of the correlation function cannot be determined
from only the \emph{correlation} matrix of the power spectrum:
the magnitude of the variances in the power spectrum,
suppressed in the fixed-amplitude mocks,
must be considered when deriving the correlation matrix of the correlation function.

\subsection{Influences on variance suppression}
\label{sec:effects}

We proceed to investigate the impact of
three-point clustering, small-scale clustering, and overall galaxy bias
on the degree of variance suppression observed in our mock catalogues
upon application of fixed-amplitude initial conditions,
and provide intuitive explanations for our observations
based on the \ezmock{} bias model.

\subsubsection{Effects of higher order clustering}

\begin{figure}
\centering
\includegraphics[width=0.8\columnwidth]{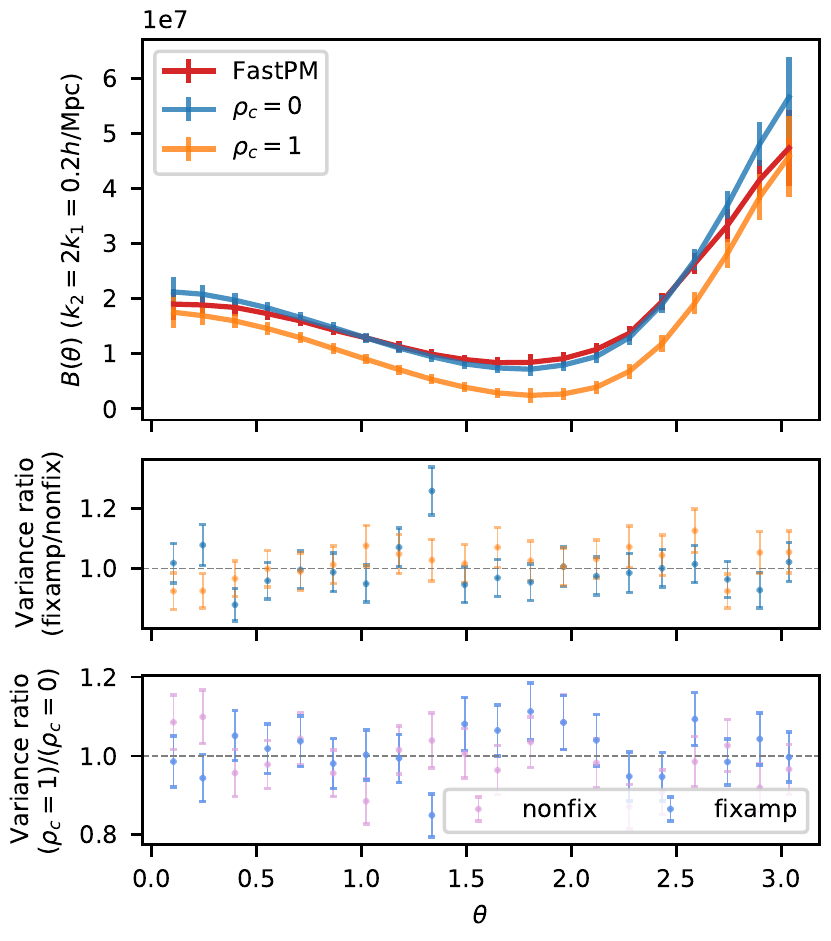}
\caption{Top: Mean ($\pm 1\sigma$) of the bispectrum~$B(\theta)$
    ($k_2 = 2k_1 = \SI{0.2}{\h\per\Mpc}$)
    for fixamp \ezmock{} catalogues
    with the same two-point clustering
    but differing density cuts~$\rho_c$,
    as described in the text.
    Middle: Ratios of $B(\theta)$ variances between fixamp catalogues
    and their nonfix counterparts
    at both values of $\rho_c$,
    demonstrating that fixing amplitudes
    yields no improvement in bispectrum uncertainty.
    Bottom: Bispectrum variance ratios between $\rho_c=1$ and $\rho_c=0$ \ezmocks{}
    for both the nonfix (pink)
    and fixamp (blue) initial conditions
    show that the variances in $B(\theta)$ are not sensitive to the mean values of $B(\theta)$.
    Standard errors computed per Appendix~\ref{sec:errorbars}.
    }
\label{fig:bk_compare}
\end{figure}

\begin{figure*}
  \centering
  \includegraphics[width=2.0\columnwidth]{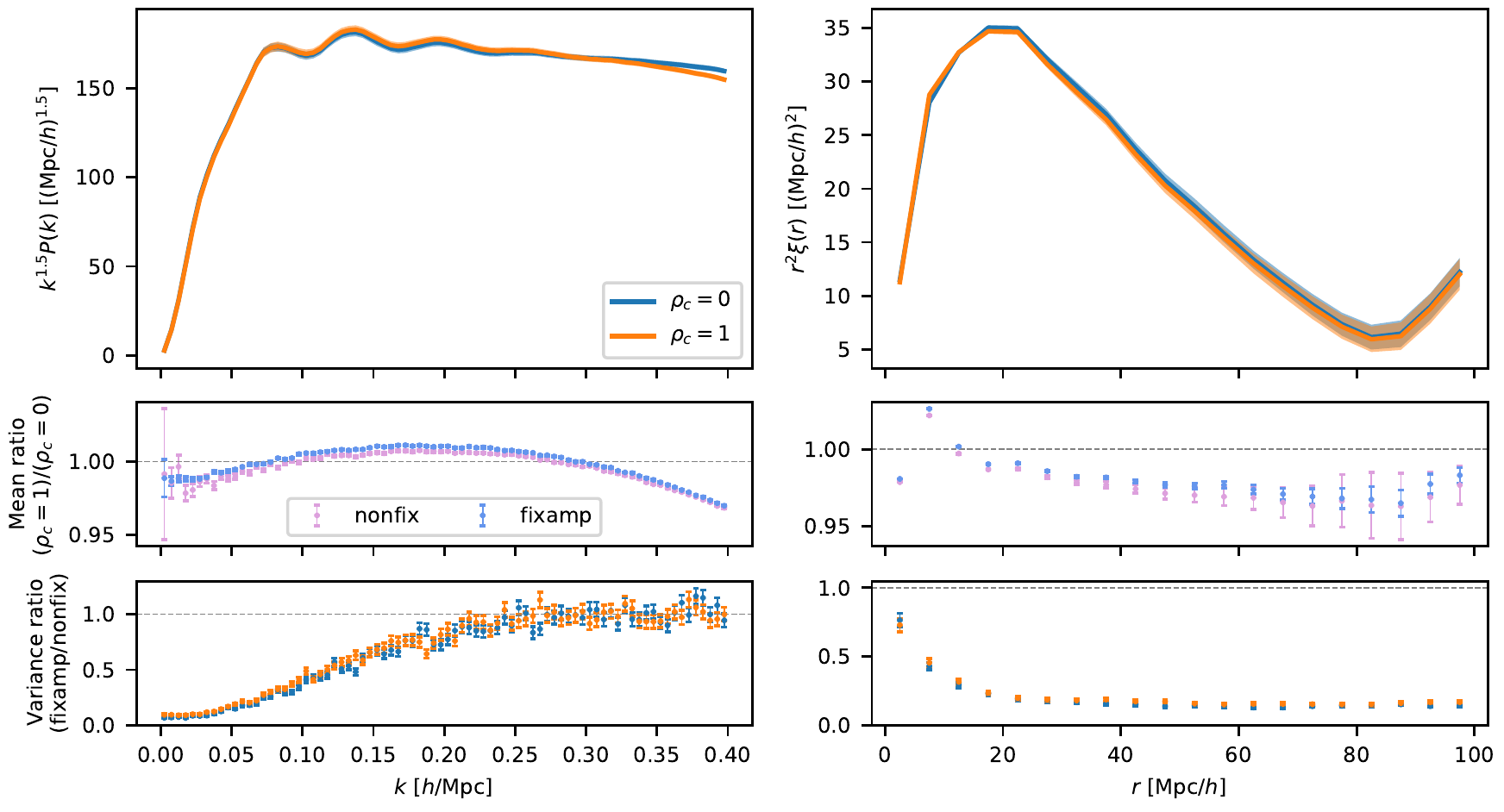}
  \caption{Comparison of two-point clustering statistics
  for two sets of 2000 \ezmocks{},
  both generated to match the \fastpm{} catalogues' two-point statistics,
  but with differing density cuts $\rho_c$ giving different three-point clustering,
  as described in the text.
  Top: Mean power spectrum and correlation function
  for the fixed-amplitude mocks in each set.
  Middle:
  Ratio of the mean values of the statistics
  (for each amplitude condition)
  between the $\rho_c = 1$ and $\rho_c = 0$ mocks showing agreement to within a few percent
  in both Fourier and configuration space.
  Bottom:
  Variance suppression obtained by the fixed-amplitude condition for both sets.
  We find slightly stronger suppression in the set with $\rho_c = 0$.
  Standard errors are computed per Appendix~\ref{sec:errorbars}.
  }
\label{fig:pk_tune_bk}
\end{figure*}

To test the impact of higher order clustering
on the variance suppression,
we compare the fiducial set of \ezmocks{}
with the second set of 2000 \ezmocks{} described in \autoref{sec:ezmock}.
The two sets have similar two-point statistics
(within 2\% for $k < \SI{0.35}{\Mpc\per\h}$),
but their differing density cuts~$\rho_c$ produce different bispectra,
as shown in the top panel of \autoref{fig:bk_compare}.
From \autoref{fig:pk_tune_bk},
the variance suppression from fixing amplitudes
appears slightly worse with greater $\rho_c$
(lower bispectrum).

One might expect this behaviour from the perspective of the \ezmock{} bias model.
\ezmock{} introduces scatter during the PDF mapping procedure
(see step~\ref{item:ezmock-pdf-mapping} of \autoref{sec:ezmock})
to reduce the clustering amplitude (galaxy linear bias):
the larger the scatter, the lower the bias.
However, when we apply a higher density cut,
we increase the linear bias of the resulting \ezmock{},
so to achieve the same two-point clustering as before,
we must then reduce the linear bias by introducing greater scatter during PDF mapping,
resulting in larger variance and weaker variance suppression.
Indeed, as previously mentioned,
the scatter procedure is the primary source of stochastic bias in \ezmock{};
it introduces variance even when we apply fixed-amplitude initial conditions.

Incidentally, the middle panel of \autoref{fig:bk_compare},
indicates that the fixed-amplitude condition
yields no significant improvement in bispectrum uncertainty,
as observed by \citet{Angulo:2016hjd}.
The lower panel of \autoref{fig:bk_compare} indicates
that the variances in the bispectrum
are not sensitive to the mean value of the bispectrum.

\subsubsection{Effects of small-scale clustering}

\begin{figure*}
    \centering
    \includegraphics[width=2.0\columnwidth]{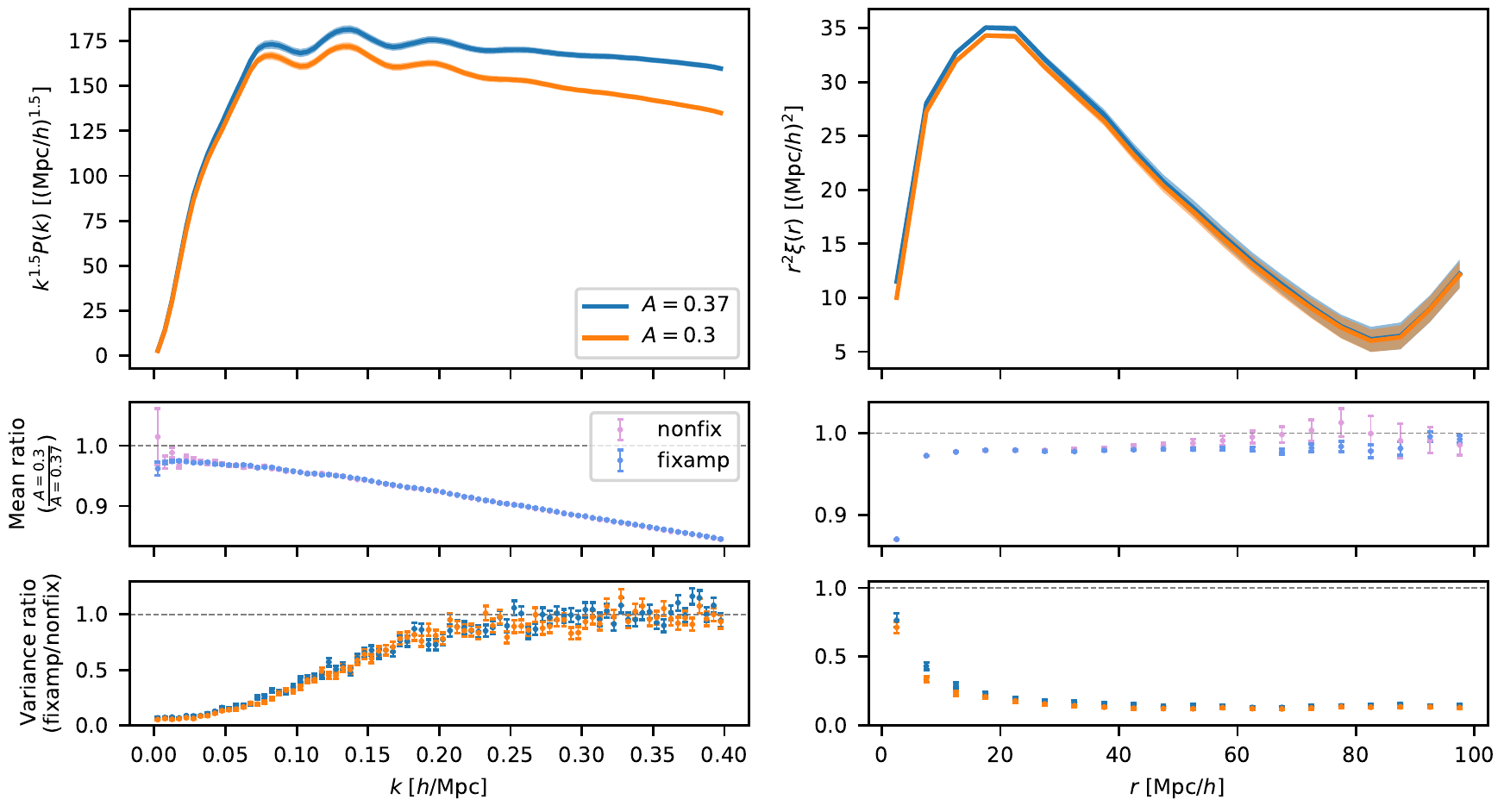}
    \caption{Two-point statistics for two sets of \ezmocks{} generated
    with different values of $A$ in \autoref{eq:pdf}
    to yield differing small-scale clustering
    but tuned to have matching large-scale clustering.
    Top: Mean statistics for the fixed-amplitude catalogues in each set (1000 each).
    Middle:
    Ratios between the mean values
    (for both initial conditions),
    showing comparable large-scale clustering
    but suppressed small-scale clustering for the $A=0.3$ mocks.
    Bottom:
    Ratios of variances in the statistics
    between the fixed-amplitude catalogues
    and their non-fixed-amplitude counterparts,
    showing better variance suppression when there is less small-scale clustering.
    Standard errors are computed per Appendix~\ref{sec:errorbars}.
    }
    \label{fig:small-scale-clustering-exp}
\end{figure*}

We next explore the effect of adjusting small-scale (large $k$) clustering.
We produced the third set of \ezmocks{} described in \autoref{sec:ezmock}
with a stronger PDF slope ($A = 0.3$ versus the original $A = 0.37$)
and thus weaker small-scale clustering
compared to the fiducial set.
In order to isolate the effect of small-scale clustering,
we adjusted the scatter parameter~$\rho_a$
to maintain the linear amplitude of the fiducial \ezmocks{}.
In \autoref{fig:small-scale-clustering-exp},
we find that these new \ezmocks{},
with weaker small-scale clustering,
show stronger variance suppression.

The \ezmock{} bias model again gives a natural explanation.
The PDF mapping procedure inherently boosts the amplitude of density fluctuations in all $k$-modes
(not necessarily by the same amount).
The scatter procedure then reduces bias at the linear scale (small $k$),
but does little to the small-scale clustering,
which remains governed by the PDF we choose (i.e.\@ $A$).
So for our new set of \ezmocks{},
with its lower small-scale clustering,
the PDF mapping procedure does not increase the linear bias
as much as it does in the fiducial set,
thus requiring less scatter to reach the targeted linear bias.
The result is the stronger variance suppression we observe.

\subsubsection{Effects of galaxy bias}

\begin{figure}
\centering
\includegraphics[width=0.9\columnwidth]{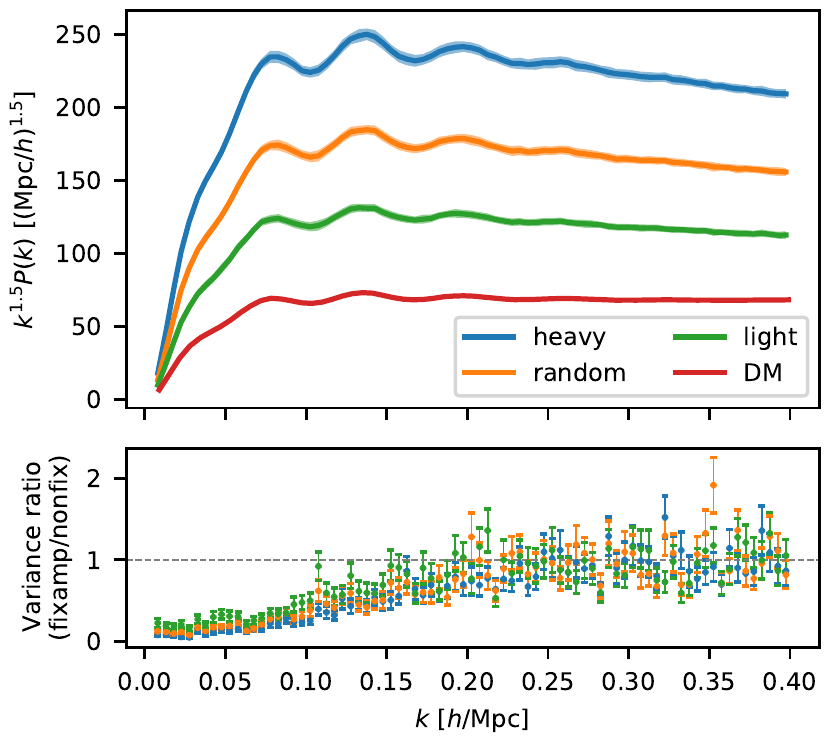}
\caption{
  Top: Power spectra from subcatalogues of the \fastpm{} catalogues.
  The subcatalogues are created in three ways:
  by selecting the top and bottom halves by mass
  (yielding ``heavy'' and ``light'' catalogues)
  and by randomly downsampling by 50\% (``random'').
  We plot mean power spectra ($\pm 1 \sigma$)
  for the subcatalogues derived from the fixed-amplitude \fastpm{} catalogues,
  as well as the dark matter power spectrum (``DM'') used in creating the \fastpm{} catalogues.
  Bottom:
  Ratios between the fixed-amplitude and non-fixed-amplitude variances,
  revealing stronger variance suppression in samples with larger bias.
  Standard errors are computed per Appendix~\ref{sec:errorbars}.
  }
\label{fig:half_catalog}
\end{figure}

Finally, we investigate how galaxy bias
affects variance suppression in fixed-amplitude catalogues.
For each of the 200 \fastpm{} catalogues,
we produce three subcatalogues
with only half the number density of the full catalogues.
The ``heavy'' and ``light'' subcatalogues
are simply the top and bottom halves of the catalogues split by mass,
while the ``random'' subcatalogue is a random 50\% downsampling.
We show the power spectra of these subcatalogues in \autoref{fig:half_catalog},
where we find that the variance improvement in the ``heavy'' subcatalogues
is noticeably stronger than in the ``light'' and ``random'' subcatalogues.
Since the subcatalogues have equal number density,
the shot-noise contributions to the variance are equal,
and we conclude that larger bias leads to stronger variance suppression.

One might expect the opposite trend:
stronger variance suppression when there is \emph{less} bias.
Indeed, we know that the underlying dark matter field
(which has unity bias, lower than any of these galaxy samples)
should show the greatest variance improvement when fixed-amplitude initial conditions are used:
there should be no dark matter power spectrum variance in fixed-amplitude simulations,
by definition.

Interestingly, the \ezmock{} bias model allows us to understand this apparent contradiction.
Indeed, \ezmock{} introduces scatter to \emph{reduce} clustering amplitudes
from some higher bias that the PDF mapping procedure would produce
in the absence of scatter.
To achieve a lower bias in the resulting \ezmock{} catalogue,
more scatter must be applied
(and thus more stochastic bias introduced).
So increased variances in clustering measurements
are associated with \emph{reduced} clustering amplitude.

\section{Conclusions}
\label{sec:conclusion}

In this study, we demonstrated that \ezmock{} with fixed-amplitude initial conditions
can be used to estimate covariance matrices for fixed-amplitude $N$-body simulations.
We further investigated the behaviour of the variance suppression
introduced by fixed-amplitude initial conditions.
Our main findings can be summarized as follows:
\begin{enumerate}
    \item No simple analytical form exists for covariance matrices of fixed-amplitude catalogues.
    \item Fixed-amplitude initial conditions do not bias clustering measurements in \ezmock{}.
    \item After calibrating \ezmock{} with clustering measurements from a reference fixed-amplitude catalogue,
    the resulting covariance matrix is a good estimate for that of the reference
    without further calibration.
    \item In fixed-amplitude catalogues, variance suppression is
        stronger with stronger three-point clustering,
        weaker when there is stronger small-scale clustering, and
        stronger when there is greater large-scale bias.
    \item The relative strength of the variance suppression in these cases
    can be understood with the \ezmock{} bias model:
    the variance suppression due to fixed-amplitude initial conditions
    is reduced when \ezmock{} requires greater scatter to reproduce
    the clustering statistics of a reference simulation.
\end{enumerate}

As survey volumes continue to grow,
variance-suppressed cosmological $N$-body simulations
will become increasingly important
in order to deliver large effective simulation volumes
at reduced computational cost.
Our work validates \ezmock{}
as an effective and efficient method for estimating covariance matrices for these variance-suppressed simulations,
paving the way for the use of the variance-suppression technique in future simulations,
and by extension,
for much more computationally cost-effective cosmological analysis.

\section*{Acknowledgements}
This work was supported in
part by U.S. Department of Energy contracts to SLAC (DE-AC02-76SF00515). 
This research has made use of NASA's Astrophysics Data System and the arXiv preprint server. \par
Some of the computing for this project was performed on the Sherlock cluster at Stanford. We would like to thank Stanford University and the Stanford Research Computing Center for providing computational resources and support that contributed to these research results.\par

\section*{Data Availability}
The data that support the findings of this study are available from the corresponding author upon reasonable request.



\bibliographystyle{mnras}
\bibliography{covariance_matrix}



\appendix
\section{Uncertainty estimation}
\label{sec:errorbars}

In this appendix,
we briefly describe
our computation of
standard errors for variances and correlation coefficients.

Indeed, as variance suppression is central to our investigation,
we must take care to compute standard errors on variances correctly.
For data $x_1, \dots, x_N$,
we estimate the variance as
\begin{equation}
    k_2 = \frac{1}{N-1} \sum_{i=1}^N (x_i - \overline{x})^2,
\end{equation}
where $\overline{x}$ is the sample mean $\frac 1N \sum_i x_i$.
This is the well-known
minimum-variance symmetric unbiased estimator of the variance.
Per \citet[p.~189]{kenney51},
an unbiased estimate of the variance of $k_2$ is given by
\begin{equation}
    \label{eq:variance-of-variance}
    \frac{2k_2^2 + \frac{N-1}{N} k_4}{N+1},
\end{equation}
where $k_4$ is the fourth $k$-statistic
\begin{equation}
    k_4 = \frac{N^2[(N+1)m_4 - 3(N-1)m_2^2]}{(N-1)(N-2)(N-3)},
\end{equation}
expressed in terms of the second and fourth sample moments about the mean~$m_{2,4}$
($m_n = \frac 1N\sum_i (x_i - \overline{x})^n $).
Taking the square root of the variance estimate \ref{eq:variance-of-variance}
yields a standard error for our variances.

We estimate uncertainties on correlation coefficients $r$
via the Fisher transformation $z = \tanh^{-1} r$,
which produces a new random variable $z$
with an approximately normal distribution
with standard deviation $\delta z \approx 1/\sqrt{N-3}$
\citep[][p.~222]{kenney51}.
To visualize any skewness in the distribution of $r$,
we plot asymmetric error bars with limits $\tanh(z \pm \delta z)$.


\bsp	
\label{lastpage}
\end{document}